# ULTRA-PERIPHERAL HEAVY ION COLLISIONS AT STAR

S. L. TIMOSHENKO* (for the STAR Collaboration)

*Moscow Engineering Physics Institute (State University), Moscow, 115409, Russia*
*E-mail: tim@intphys3.mephi.ru

Ultra-peripheral heavy ion collisions involve long range electromagnetic interactions at impact parameters larger than twice the nuclear radius, where the strong nucleon-nuecleon interactions are uneffective. We present recent results from the STAR collaboration on these ultra-peripheral interactions likes incoherent $\rho^0$ production in deuteron-gold collisions and interferometry in gold-gold collisions.

## 1. Introduction

In ultra-peripheral heavy-ion collisions, two colliding nuclei geometrically miss each other. The impact parameter $b$ is larger than twice the nuclear radius $R_A$ and at such conditions hadronic interactions do not occur [1]. Relativistic heavy ions are a powerful source of coherent electromagnetic and Pomeron fields. The Pomeron carries the strong interaction but is colorless and it has the quantum numbers of the vacuum [2]. The electromagnetic fields have a larger radius of interaction than strong interactions. So, in ultra-peripheral heavy ion collisions, the nuclei primarily interact by two photons or by photon-Pomeron exchange. Here we will present recent data on vector meson production and vector meson interferometry which has been obtained by the STAR collaboration. The STAR collaboration has already published results on $\rho^0$ production in $AuAu$ collisions at an energy of 130 GeV per nucleon [3].

Exclusive vector meson production $dAu \to d(np)Au\rho^0$ (fig.1(a,b)) can be described by using the Weizsacker-Williams approach to photon fluxes [4] and the vector meson dominance model [5]. One can consider the photon emitted by one nucleus as a state of virtual photons plus some fluctuations of $q\bar{q}$ pairs. When the nucleus absorbs the "photonic" part of wave function, the $q\bar{q}$ pairs contribution becomes dominant. This pair can elastically scatter on the other nucleus and appear as a real vector meson. The "Pomeron" represents the absorptive part of nuclear cross section.

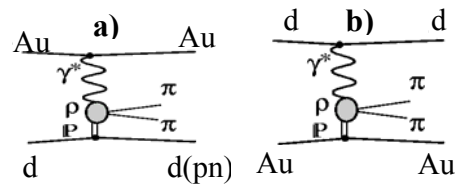

Fig.1 Diagrams for exclusive $\rho^0$ production in ultra-peripheral deuteron gold collisions with photon emitted by gold (a) and deuteron (b).

The photon flux is proportional to the square of the nuclear charge $Z^2$. The forward cross section scales as $A^2$ for heavy mesons, while a Glauber calculation finds that shadowing reduces the dependence for the $\rho^0$ to approximately $A^{5/3}$ [5]. Thus the cross section for photon-Pomeron interaction goes as $Z^2 A^{5/3}$ for $\rho^0$ meson.

In ultra-peripheral deuteron gold interactions there are two mechanisms of $\rho^0$ meson production. First, a photon emitted by the gold nucleus interacts with the deuteron and produces $\rho^0$ meson. In this case, deuteron can break up ($\gamma d \to np\rho^0$) or remain in the ground state ($\gamma d \to d\rho^0$) (Fig. 1a). But there is also process in which a photon emitted by a deuteron interacts with the gold nucleus (Fig. 1b). In $dAu$ collisions the photon usually comes from gold nucleus, removing the photon direction ambiguity present in the gold-gold collisions.

The STAR collaboration has also studied 4-prong final states and recently presented the first preliminary data on coherent production of $\pi^+\pi^-\pi^+\pi^-$ [6].

## 2. The STAR Detector

In the years 2001 and 2003, the RHIC accelerator at Brookhaven National Laboratory operated in gold-gold (2001) and deuteron-gold (2003) modes at $\sqrt{s_{NN}} = 200$ GeV. In the Solenoidal Tracker at RHIC (STAR), charged particles are reconstructed in a 4.2 meter long, 4 meter diameter cylindrical time projection chamber (TPC) [7]. A solenoidal magnet surrounds the TPC. The TPC was operated in a 0.5 T solenoidal magnetic field. Particles were identified by their energy loss in the TPC. The reconstruction efficiency for charged pions is high for tracks with transverse momentum $p_T > 100$ MeV/c and pseudorapidity $|\eta|<1.15$. The TPC is surrounded by a cylindrical central trigger barrel (CTB). The CTB consists of 240 scintillator slats covering $|\eta|<1$. There are two zero degree calorimeters (ZDC) at $\pm 18$ m from the interaction point for neutron detection [8]. These ZDCs are sensitive to single neutrons and have efficiency close to 100%.

## 3. $\rho^0$ production

Exclusive $\rho^0$ production in UPC has a distinctive experimental signature: the $\pi^+\pi^-$ decay products of the $\rho^0$ meson are observed in an otherwise 'empty' spectrometer. Two different triggers are used for this analysis. For $dAu \to dAu\rho^0$, about 700 thousand events were collected using a low-multiplicity 'topology' trigger. The topology trigger was designed to detect the products of $\rho^0$ decay in the CTB system. The CTB was divided into four azimuthal quadrants. Event with hits in the north and south sectors were selected. The top and bottom quadrants were used for the rejection of cosmic rays. To study $dAu \to npAu\rho^0$, about 250 thousand events were collected using a 'topology-ZDC' trigger. This trigger required detection of neutrons from deuteron break up as well as hits in the North and South CTB quadrants. The main backgrounds for two triggers are cosmic rays, beam gas interaction, and pile-up.

This analysis selected events with exactly two reconstructed tracks in the TPC. Events were accepted if two oppositely charged tracks formed a common vertex within the interaction region.

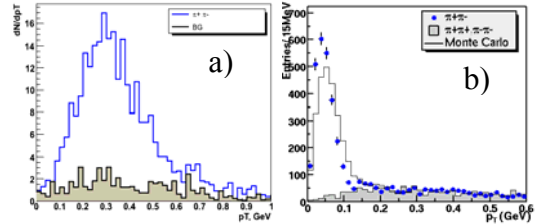

Fig. 2. The $\rho^0$ transverse moment in deuteron gold (a) and gold-gold (b) collisions, the shaded histograms are combinatoric hadronic background..

Figure 2a shows the transverse momentum spectrum of oppositely charged pion pair production in deuteron gold collisions. One can see a large peak in the region of 300 MeV. The $dAu$ results can be compared with $\rho^0$ meson production in gold-gold collisions [9]. Fig. 2b shows the transverse momentum spectrum for $\pi^+\pi^-$ pairs in $AuAu$ collisions. A clear peak, the signature for coherent coupling, can be observed at $p_T < 100$ MeV. This is consistent with coherent $\rho^0$ meson production. A background model for like-sign combination pairs, which is normalized to the signal at $p_T > 250$ MeV, does not show such a peak. The broader peak in the $p_T$ spectrum in $dAu$ collisions corresponds to photon-deuteron interactions, with no indication of photon-gold interactions, as expected here, since the deuteron breaks up. The shape of the signal distribution is sufficiently different from the background distribution that the

background cannot be normalized to agree with the signal curve over the full range in $p_T$

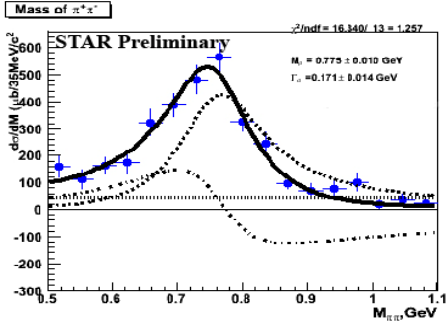

Fig. 3. The $d\sigma/dM_{\pi\pi}$ invariant mass distribution for 2-track events.

Fig. 3 shows the invariant mass of the $\pi^+\pi^-$ pairs for "topology-ZDC" triggers. One can see a clear $\rho^0$ peak in this distribution. The fit (solid) is the sum of a relativistic Breit-Wigner for $\rho^0$ production and a Soding's interference term [5] for direct $\pi^+\pi^-$ production (both dashed). The interference shifts the $\pi^+\pi^-$ distribution to lower $M_{\pi^+\pi^-}$. The $\rho^0$ mass and width are consistent with accepted values [10]. The direct $\pi^+\pi^-$ to $\rho^0$ ratio agrees with studies by the ZEUS collaboration in $\gamma p$ interactions [11].

We study the $p_T$ spectra using the variable $t_\perp = p_T^2$. At RHIC energies, the longitudinal component of the 4-momentum transfer is small, so $t \approx t_\perp$. Figure 4a shows a $t$ spectrum. The $t$ spectrum drops off at small t. We have compared our data with the data of fixed target ($\gamma d$) (Fig. 4b) [12]. One can see that the incoherent $t$-behavior is similar to the $t$ distribution in Fig.4a. One can conclude that the $\rho^0$ meson production in reaction, when photon interacts with deuteron and deuteron break up. The data were fitted with the function $dN/dt \sim \exp(-bt)$. We obtained $b = 9.2 \pm 0.2$ GeV$^{-2}$ and it agrees with data the $b$ measured in the fixed target experiments [12].

## 4. $\rho^0$ interferometry

In ultra-peripheral heavy ion collisions, a nucleus can be either the photon source or the scattering target. In vector meson photoproduction these two possibilities are indistinguishable, and they should be able to interfere. Since the $\rho^0$ have negative parity, the interference is destructive. So the two amplitudes subtract with a transverse momentum ($p_T$) dependent phase factor to account for the separation. The cross section is [13]

$$\sigma = |A_1 - A_2 \exp(ip_T \cdot b)|^2 \qquad (1)$$

where $A_1$ and $A_2$ are the amplitudes for $\rho^0$ production from the two directions. At mid-rapidity $A_1 = A_2$ and

$$\sigma = \sigma(b)[1 - \cos(p_T \cdot b)] \qquad (2)$$

where $\sigma_0(b)$ is the cross section without interference. The system acts as a 2-slit interferometer, with slit separation $b$ [14]. The pt spectrum is suppressed for $p_T < \hbar/\langle b \rangle$, where $\langle b \rangle$ is the median impact parameter.

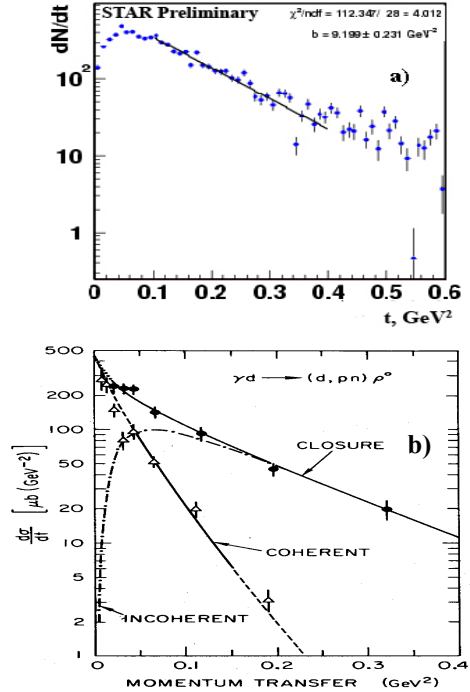

Fig. 4. The differential cross section $d\sigma(\gamma d \to \rho^0 np)/dt$ for the same data set (a) and $d\sigma(\gamma d \to \rho^0 np)/dt$ (b) [12].

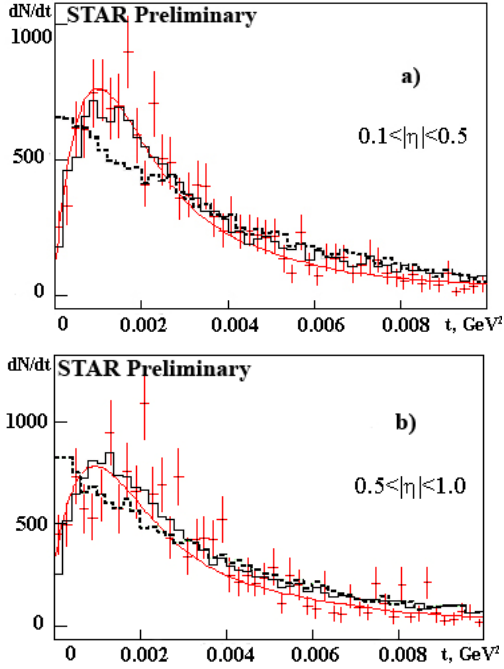

Fig. 5. Raw (uncorrected) $t_\perp$ spectrum for $\rho^0$ sample for $0.1 < |\eta| < 0.5$ (a) and $0.5 < |\eta| < 1.0$ (b). The dashed and solid histograms represent simulations without and with interference; the solid line is the result of the fit to Eq. 3.

Figure 5 compares the uncorrected minimum bias data for $0.1 < |\eta| < 0.5$ and $0.5 < |\eta| < 1.0$ with simulations with and without interference. Simulations include the detector response. The data drops off for t < 0.001 GeV$^2$. This drop is due to interference. The data is fit to the 3-parameter form:

$$\frac{dN}{dt} = a\exp(-bt)[1 + c(R(t) - 1]  \quad (3)$$

where $R(t) = Int(t)/Noint(t)$ is the ratio of the Monte Carlo $t$-spectra with and without interference. Here, a is the overall normalization, the slope $b \approx R_A^2$, and $c$ is the degree of spectral modification; $c = 0$ corresponds to no interference while $c = 1$ is the calculated interference. This functional form separates the interference ($c$) from the nuclear form factor ($b$).

For this data set, the interference is $93 \pm 6(stat.) \pm 8(syst.) \pm 15(theory)\%$ of that expected [14]. Because the two sources are spatially separated, the final state $\pi^+\pi^-$ wave function does not factorize into single-particle wave functions, and the system exhibits the Einstein-Podolsky-Rosen paradox [15].